# FORMAL VERIFICATION OF SPACE SYSTEMS DESIGNED WITH TASTE

I. Dragomir [(1)], M. Bozga [(2)], I. Ober [(3)], D. Silveira [(4)], T. Jorge [(4)], E. Alaña [(1)], M. Perrotin [(5)]

[(1)] GMV Aerospace and Defence, Isaac Newton 11, PTM, Tres Cantos, 28760, Spain, Email: {idragomir, ealana} @gmv.com

[(2)] VERIMAG, CNRS, 700 Av Centrale, 38401 St. Martin d'Hères, France, Email: marius.bozga@univ-grenoble-alpes.fr

[(3)] ISAE-Supaero, Univ. of Toulouse, 10, avenue Édouard-Belin, 31055 Toulouse, Email: iulian.ober@isae.fr

[(4)] GMVIS Skysoft SA, Alameda dos Oceanos, 115, Lisbon, 1990-392, Portugal, Email: {daniel.silveira, tmdasilva} @gmv.com

[(5)] European Space Agency (ESA), ESTEC, Keplerlaan 1, PO Box 299, Noordwijk, The Nertherlands, Email: maxime.perrotin@esa.int

**ABSTRACT**

Model-Based Systems Engineering (MBSE) is a development approach aiming to build correct-by-construction systems, provided the use of clear, unambiguous and complete models to describe them along the design process. The approach is supported by several engineering tools that automate the development steps, for example the production of code, documentation, test cases and more.

TASTE [1] is pragmatic MBSE toolset supported by ESA that encapsulates several technologies to design a system (data modelling, architecture modelling, behaviour modelling/implementation), to automatically generate the binary application(s), and to validate it. One topic left open in TASTE is the formal verification of a system design with respect to specified properties. In this paper we describe our approach based on the IF model-checker [4] to enable the formal verification of properties on TASTE designs. The approach is currently under development in the ESA MoC4Space project.

## 1. INTRODUCTION

Developing space systems is a difficult task, where the engineers need to address two challenges: (1) how to design large and complex systems with minimal effort and cost, and (2) how to ensure that the designed system is correct with respect to its requirements. Model-Based Systems Engineering (MBSE) is an adopted development approach aiming to address the above challenges by enabling correct-by-construction system design from which the implementation could be generated and deployed on the target platforms.

More specifically, MBSE allows to produce unambiguous, consistent and coherent designs of the system which can be subject to validation and verification with respect to system properties during the development cycle. By design we understand system architecture, behaviour and data types, as well as deployment strategy. By validation and verification we understand an assortment of techniques including design review, testing, interactive simulation and model-checking. While the first three methods allow to detect errors within the design in a lightweight manner (i.e., only a subset of the system's behaviours are evaluated), model-checking exhaustively analyses the system's behaviour and gives a verdict for property satisfaction.

TASTE [1] is an MBSE toolset supported by ESA which enables system design and validation. Several languages are available for system design: ASN.1 for data types, a flavour of AADL for hierarchical architecture, SDL for behaviour modelling, and Ada/C/C++ when manual coding is needed. The toolset offers as features static type analysis, real-time scheduling analysis, code generation and binary applications generation for target platforms, as well as simulation, debugging and testing. A feature not yet supported by the toolset is the formal verification of the system design with respect to system properties.

The ESA *Model-Checking for Formal Verification of Space Systems* (MoC4Space) project[1] aims to develop and integrate in TASTE various model-checking tools to automate the formal verification of system properties. In the following we describe the approach considered and the development status.

## 2. MODEL-CHECKING TASTE DESIGNS

### 2.1. Overview

The proposed model-checking approach for a TASTE design and related properties, consists of the following steps as illustrated in the workflow from Figure 1:

1) The user designs the system with TASTE. The design consists of data view (ASN.1 data types), interface view (AADL hierarchical system architecture), SDL state machines/C implementation/GUI function (system behaviour).

2) The user specifies with TASTE the safety properties the system shall fulfil. Properties can be specified in three

---

[1] This project is funded by the European Space Agency under contract number 4000133658/21/NL/CRS.

ways, from the least to the most expressive: Boolean stop conditions, Message Sequence Chart properties, and dynamic observer properties.

3) The user invokes the model-checker. The user selects the properties to be checked from those specified and sets the configuration of the model-checker. The model-checker works on the system design and the specified properties. The system behaviour is explored and the properties satisfaction is checked on all possible behaviours. A result is provided to the user: either the properties are validated and a message is displayed, or a property is violated and diagnostic traces are produced.

4) The user checks the result provided by the model-checker. If the properties are satisfied, then the workflow stops. If the property is not satisfied, the obtained diagnostic traces are played in TASTE and assessed individually. If the diagnostic trace shows a valid behaviour for the system, then the defect is corrected and the process starts again. If the diagnostic trace shows an invalid behaviour, the next available diagnostic trace is analysed or the process stops. (Note that the model-checker will intentionally explore a superset of the system's executions, in particular by relaxing certain scheduling constraints. In consequence, diagnostic traces can show invalid behaviours.)

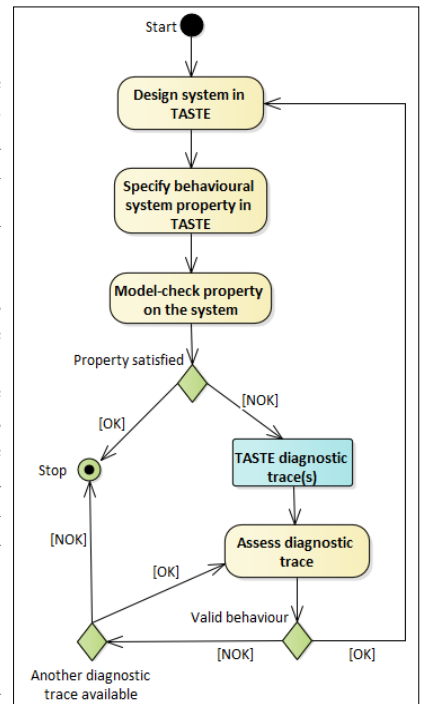

## 2.2. Tool Design

To enable the workflow explained above, the following extensions are required: (1) property specification in TASTE, (2) model-checker for TASTE designs and properties and (3) seamless integration of the approach in the TASTE GUI.

2.2.1. Property Specification in TASTE

Three types of properties are considered for model-checking: Boolean stop conditions (BSC), Message Sequence Charts (MSC), and dynamic observers.

Boolean stop conditions describe invalid behaviour of the system, e.g., *stop if (battery_level < 10)*. These properties are Boolean expressions that express system invariants (through their negation) and that should not evaluate to *true* at any point during the execution. Informally, the satisfaction of the Boolean condition means that an undesired state has been reached, and therefore the system design is not correct. Such properties are modelled in TASTE as dynamic observers in OpenGEODE (see below). The user will select *BSC property* from a dedicated menu in the TASTE GUI, and an SDL observer (with 2 states and 1 decision) is generated and displayed using OpenGEODE. Then the user can directly specify the condition with the available SDL syntax for observers.

MSC properties describe the (un)desired behaviour of the system as a sequence of I/O events happening between the system's functions. Such properties are specified using the MSC editor available in TASTE. The user will select *MSC property* from the dedicated TASTE GUI menu, and the MSC editor is opened for the user to specify the property. The meaning of an MSC as a system property depends of specific annotations: *property type: search [[from-start|nonstrict]] ((intended|unintended))* or *property type: verify [[from-start]] intended*, where [[ ]] denotes an optional part and | denotes a choice. *Search* indicates that the model-checker will look for a system execution complying with the MSC. *Verify* indicates that all system executions must comply with (a prefix of) the MSC. *From-start* indicates that the MSC shall be matched from the beginning of the execution between the specified functions. *Nonstrict* indicates that other I/O events that are not specified can happen between the functions. *Intended* indicates that the MSC is a desired execution sequence, while *unintended* indicates that the MSC is an undesired sequence. Then the model-checker will provide a result consistently with the type of property indicated by the user.

Dynamic observers describe both desired and undesired behaviour of the system in the form of state machines. They are the most expressive properties considered, as they can monitor the system execution, but also alter it for example to guide the verification process. Observers are represented as SDL state machines in OpenGEODE. The user will select the *Observer property* in the dedicated TASTE GUI menu, and OpenGEODE will be displayed in which the property is modelled. The user has then access to the SDL extensions for observers that allow them to observe (1) the state of the TASTE system at a point in its execution (specifically, the values of variables and state machine location for the functions) and (2) the events that occurred during the last atomic transition taken by one of the SDL functions of the TASTE system. The user specifies whether the execution is desired or undesired by labelling the states of the observer with *success* or *error* accordingly. Then the model-checker will correctly assess the property on the system executions.

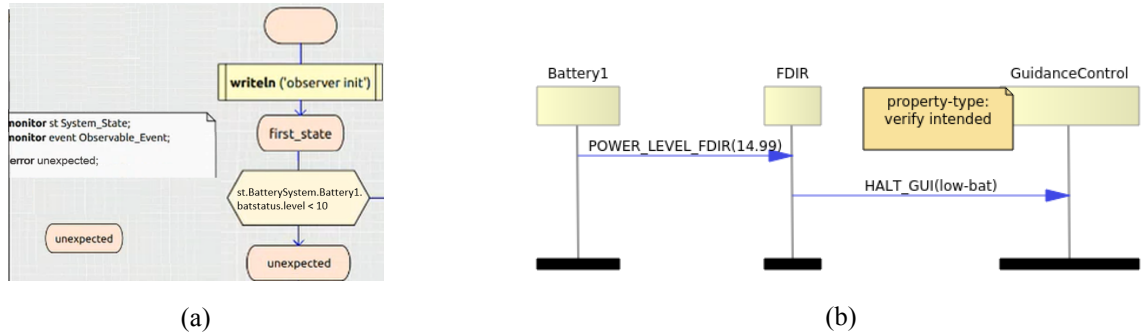

(a)    (b)

*Figure 2. Examples of properties with TASTE: (a) Boolean stop condition / dynamic observer modelling an error if the battery level is lower than 10 units, (b) MSC property modelling that if the battery level drops below 15 the FDIR stops the system.*

2.2.2.  Integrated Model-Checking

Model-checking ([2], [3]) is a well-known formal verification technique for the system correctness with respect to a set of properties. It consists in building a finite state-space model of the system under analysis and checking the properties on this model. The check itself amounts to a complete or partial exploration of the state space, potentially guided by the properties. The main advantage of model-checking is that it can be fully automated. Moreover, it allows for the production of diagnostic traces when a property is not verified, that can help the engineer to correct the design. The main drawback is, however, the potential size of the state-space model which depends on the system complexity and which could make model-checking infeasible.

Several solutions can be considered for developing a model-checker for TASTE: (1) re-use of an existing off-the-shelf model-checker (e.g., IF [4], UPPAAL [5]), (2) re-use an existing open model checker platform (e.g., LTSmin [6]) or (3) develop a dedicated model-checker. From the trade-off analysis of the different options, we have selected to re-use the IF toolset as model-checking back-end.

The IF toolset [4] is an environment for modelling and validating real-time systems, characterised by the following key features: (1) a modelling formalism (also called IF) based on extended Timed Automata with urgencies and asynchronous FIFO-based communication, syntactically and semantically close to SDL, (2) properties specification to be validated on models as dynamic observers, (3) combined use of various validation techniques including model-checking, static analysis of the IF model representation and simulation, (4) generation of diagnostic traces in the case of property violations, and (5) support for high level modelling with formalisms such as SDL, UML, SysML used in industrial CASE tools, through translation of high level models into the IF notation.

Using the IF model-checker as back-end requires that bidirectional transformation are defined between TASTE and IF. These transformations cover (1) generating the IF system model from the TASTE design views, (2) generating the IF observers from the TASTE system properties, and (3) generating TASTE MSC representation from the model-checker diagnostic traces. An overview of the tool architecture in provided in Figure 3.

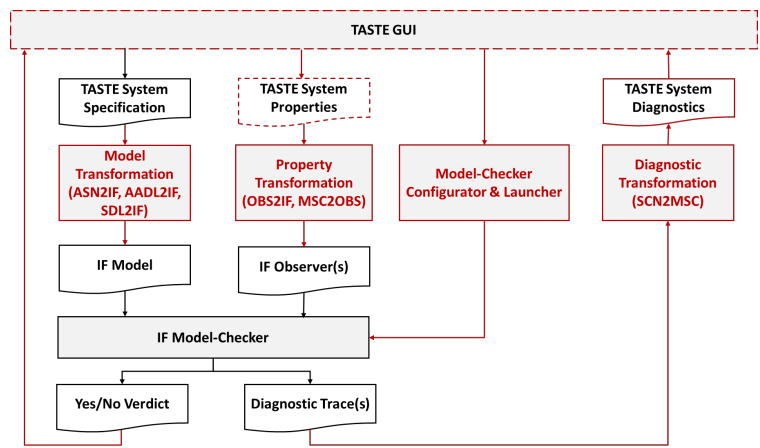

*Figure 3. TASTE model-checking tool architecture. Boxes in red represent new/enhanced modules, black arrows existing information/ dependency flow, and red arrows new information/dependency flow.*

Model transformation implies obtaining an IF representation from the TASTE design. This module is composed of several components. ASN2IF transforms the ASN.1 datatypes specified in the TASTE data view to IF datatypes. AADL2IF transforms the system structure specified in the TASTE interface view to an IF model. This component also handles the specific transformation for GUI functions (modelling the environment), C functions behaviour and timers specified in the interface view. SDL2IF transforms the behaviour of SDL functions modelled as state machines to IF processes.

Property transformation implies obtaining an IF observer for each TASTE system property. A dynamic SDL observer is transformed to an IF observer (OBS2IF) in mostly the same way as an SDL state machine is transformed to an IF process. In fact, the OBS2IF component is an extension of SDL2IF component with the specific handling of the observers' syntax and semantics. A BSC property is modelled as an SDL observer, so the same OBS2IF component is used for obtaining an IF observer from them. Finally, an MSC property is also transformed into an SDL observer (MSC2OBS). This transformation has the benefit of obtaining and displaying graphically the property to be checked into another formalism known by the user, i.e., SDL. Also, the user has the possibility of enhancing the property represented as SDL observer as desired, since observers are more expressive than MSC properties. Then the OBS2IF is called on the SDL observer representation of the MSC to obtain the corresponding IF property that will be checked.

Diagnostic transformation implies obtaining an MSC from the diagnostic traces provided by IF (SCN2MSC). The IF model-checker provides diagnostic traces as a sequence of IF transitions steps, each including labels of the actions executed (inputs, outputs, process creation, deletion, etc.). MSC is the most usable formalism to visualize such diagnostic traces and to replay them in the TASTE environment, thus easing the understanding of the result(s).

To provide a seamless integration in TASTE of the model-checker, the TASTE GUI is extended with a model-checker configurator and launcher (see Figure 3). The configurator allows to specify the environment of the system under validation, the properties to be checked and the model-checker options (e.g., number of diagnostic traces generated, number of states explored), The launcher allows to call and stop the IF model-checker. The TASTE build system is also extended to invoke the above-mentioned transformations and model-checker.

### 2.3. Tool Validation

The TASTE model checking approach is validated on two industrial case studies: the IXV mission and the ERGO planetary scenario [7]. The IXV mission aimed to define the basic needs for re-entry from Low Earth Orbit. The case study considered in this activity is a subset of the fully-automated on-board software that focuses in the flaps control system. The ERGO planetary scenario is inspired by the Mars Sample Return (MSR) mission that covers the concepts and requirements of the Martian Long Range Autonomous Scientist. The case study considered in this activity consists of a subset of functionalities, "simulating" (simplified) traverse and sample collection, image acquisition, as well as system commanding in E1 (telecommanding) and E4 (goal commanding) autonomy modes. Both case studies have been modelled in TASTE with the concepts supported by the model-checker and several properties of each type have been defined for both (the complete TASTE design is available at [8]). The model-checking approach will be used to check the correctness of the modelled systems with respect to their properties. Also, errors are specifically introduced in the system design to validate the model-checking results.

### 3. DISCUSSION

The MoC4Space project aims to develop formal verification system designs and properties modelled with TASTE. The model-checking approach has been defined, as well as the tool architecture based on the existing IF model-checker as backend. The implementation of the integrated model-checking tool (property modelling editors, ASN2IF and SDL2IF) is ongoing. The approach and tool are validated and will be used to verify two space systems case studies fully developed with TASTE for the purpose of the activity.

### 4. REFERENCES

1. TASTE: The Assert Set of Tools for Engineering. Available at: https://taste.tools/
2. Queille, J. P.; Sifakis, J. (1982), "Specification and verification of concurrent systems in CESAR", *International Symposium on Programming*, Lecture Notes in Computer Science, 137: 337–351, doi:10.1007/3-540-11494-7_22, ISBN 978-3-540-11494-9
3. Clarke, E. M.; Emerson, E. A.; Sistla, A. P. (1986). *Automatic verification of finite-state concurrent systems using temporal logic specifications*, ACM Transactions on Programming Languages and Systems, 8 (2): 244, doi:10.1145/5397.5399
4. The IF toolset. Available at http://www-verimag.imag.fr/~async/IF/
5. UPPAAL (Available at http://www.uppaal.org)
6. LTSmin (Available at https://ltsmin.utwente.nl)
7. European Robotic Goal-Oriented Autonomous Controller (ERGO) H2020 project. https://www.h2020-ergo.eu/
8. MoC4Space GitLab. https://gitrepos.estec.esa.int/taste/if-model-checking.git